\documentclass[twocolumn]{aastex631}
\usepackage{showyourwork}
\usepackage{graphicx}
\usepackage[ruled,vlined,linesnumbered]{algorithm2e}
\usepackage{algorithmic}

\SetKwInput{Parameters}{Parameters}
\SetKwInput{Variables}{Variables}

\usepackage{tabularx,booktabs,multirow,amsmath}
\usepackage{float}
\usepackage[italicdiff]{physics}

\newcommand{\ripple}{\texttt{ripple}}

\newcommand{\jax}{\texttt{jax}}
\newcommand{\zdethp}{\texttt{zdethp}}

\definecolor{rb4}{HTML}{27408B}

\definecolor{cyan}{HTML}{0097A7}

\definecolor{rr}{RGB}{173, 37, 37}

\newcommand{\cuhk}{\affiliation{Department of Physics, The Chinese University of Hong Kong, Shatin, N.T., Hong Kong}}
\newcommand{\flatiron}{\affiliation{Center for Computational Astrophysics, Flatiron Institute, New York, NY 10010, USA}}
\newcommand{\JHU}{\affiliation{William H. Miller III Department of Physics and Astronomy, Johns Hopkins University, Baltimore, Maryland 21218, USA}} 

\begin{document}

\title{Recalibrating Gravitational Wave Phenomenological Waveform Model}

\author{Kelvin K.~H.~Lam} 
\email{kelvin33550336@gmail.com}
\cuhk
\author{Kaze W.~K.~Wong} 
\flatiron
\author{Thomas D.~P.~Edwards}
\JHU

\begin{abstract}
	We investigate the possibility of improving the accuracy of the
	phenomenological waveform model, IMRPhenomD, by jointly optimizing all the
	calibration coefficients at once, given a set of numerical relativity (NR)
	waveforms. When IMRPhenomD was first calibrated to NR waveforms, different
	parts (i.e., the inspiral, merger, and ringdown) of the waveform were calibrated separately. 
	Using \ripple, a library of waveform models compatible with automatic differentiation, we can, for the first time,
	perform gradient-based optimization on all the waveform coefficients at the same time.
	This joint optimization process allows us to capture previously ignored correlations between separate parts of the waveform.
	We found that after recalibration, the median mismatch between the model and NR waveforms decreases by $50\%$.
	We further explore how different regions of the source parameter space respond
	to the optimization procedure. We find that the degree of improvement
	correlates with the spins of the source. This work shows a promising avenue
	to help understand and treat systematic error in waveform models.
\end{abstract}

\section{Introduction} \label{sec:intro}

Many data analysis tasks in gravitational wave (GW) astrophysics, such
as match filtering~\citep{owen1996search, owen1999matched} and parameter 
estimation~\citep{Dax:2021tsq, Christensen:2022bxb, dynesty, Islam:2022afg, 
Romero-Shaw:2020owr, zackay2018relative, Ashton:2018jfp}, rely upon accurate waveform
models. Since the generation of numerical relativity (NR) waveforms is
prohibitively expensive, the community has constructed waveform approximants
that can be evaluated much faster. There are three families of commonly used GW
approximants: the effective-one-body (EOB) \citep{ossokine2020multipolar,
cotesta2020frequency, taracchini2014effective}, NR
surrogate \citep{islam2022surrogate, varma2019surrogate, varma2019surrogate2},
and phenomenological (Phenom) models \citep{husa2016frequency,
khan2016frequency, garcia2020multimode, pratten2021computationally}. While the
detailed construction of each model is different, they all have a set of
internal parameters that can be calibrated to NR waveforms. The quality of the
waveform model is therefore determined by the ansatz used and the accuracy of the
calibrated parameters.

The LIGO-VIRGO-KAGRA (LVK) collaboration
~\citep{LIGOScientific:2014pky,LIGOScientific:2021usb,LIGOScientific:2021djp,
VIRGO:2014yos,KAGRA:2020tym} recently started their fourth observational
run on May 26, 2023. Impressively, they are
expected to double the total number of observed binary black holes (BBHs)~\citep{abbott2020prospects}.
Moreover, the improved sensitivity also implies that we expect to
detect individual events with a higher signal-to-noise ratio (SNR) than ever before. This means
we can resolve more features in the signal, therefore putting more
stringent requirements on the accuracy of our waveform model \citep{purrer2020gravitational, hu2022assessing}.

Because of the large number of calibration parameters (often a few hundred if not
more), waveform models are usually calibrated separately for the inspiral, merger,
and ringdown parts of the waveform \citep{khan2016frequency, 
santamaria2010matching, pratten2021computationally}. This
ignores the correlation between different parts of the waveform model and limits
it quality. Recently, there has been an effort to rebuild waveform models
\citep{khan2016frequency} using programming languages that support automatic differentiation (AD) 
\citep{ripple, Iacovelli:2022bbs, Iacovelli:2022mbg, Coogan:2022qxs}; AD is a technique 
used to compute machine precision derivatives of functions 
without the issues of scaling up to high dimension or expression swelling. In
particular, \ripple~\citep{ripple} exposes the calibration parameters to the
user. This allows us to make use of common techniques from
machine learning, such as gradient descent and back propagation \citep{jax2018github, 
pytorch, tensorflow2015}, to improve the calibration of the waveform models.

In this paper, we investigate the possibility of further improving the accuracy
of a waveform model, IMRPhenomD \citep{khan2016frequency, husa2016frequency}, by jointly optimizing all the
calibration coefficients for given a set of NR waveforms. Using a similar set of NR
waveforms to those used in \citep{khan2016frequency, husa2016frequency},
we demonstrate that one can improve the match between IMRPhenomD and NR
waveforms over a decently sized parameter space, up to a mass ratio $q=8$. We additionally
explore how different parts of the source parameter space (e.g. the primary and
secondary spins) respond to the optimization procedure by optimizing the waveform
separately for different regions. This can help in understanding
whether the waveform model ansatz performs equally well in different regions of the
parameter space.

The rest of the paper is structured as follows: In Sec.~\ref{sec:method}, we
review the parameterization of the IMRPhenomD model and the mismatch function
that is used as an objective/loss function for the calibration, followed by outlining the
specific optimization scheme used for recalibration. In
Sec.~\ref{sec:result}, we give the optimization result by comparing mismatches
of the optimized waveforms with the original waveforms. We also show how the
optimization result differs as a function of the source parameters. Finally, in
Sec.~\ref{sec:discussion}, we discuss the difference between our calibration
procedure and the procedure used in \citep{khan2016frequency}. We also explain how the reduced spin
parameterization affects the accuracy of the model. 
Note that throughout this paper we use the terms recalibration and optimization interchangeably.

\section{Optimization Method} \label{sec:method}

In this section, we first briefly review the construction of the IMRPhenomD model and discuss
how the calibration parameters enter the waveform.
We then outline the mismatch and how it can be used as a loss function. 
Finally we discuss the gradient descent algorithm and our stopping criterion.

\subsection{Waveform Model} \label{subsec:waveform_model}

We start by giving a succinct summary of the IMRPhenomD model and the relevant parameters.
Interested readers should refer to \citep{khan2016frequency} for more details.

Aligned-spin, frequency-domain waveform models (such as IMRPhenomD) can be written as a
combination of amplitude and phase functions ($A$ and $\phi$ respectively):
\begin{align}\label{eq:}
	h(f,\theta,\Lambda) = A(f,\theta,\Lambda)e^{-i\phi(f,\theta,\Lambda)}\,,
\end{align}
where $f$ is the frequency, $\theta$ are the intrinsic parameters of the binary, and $\Lambda$ is a set of additional
parameters which will be discussed below. 
The phase and amplitude functions are then split into three sections which represent the
inspiral, intermediate, and merger-ringdown (MR) parts of the waveform. 
During inspiral, $A$ and $\phi$ are known analytically from post-Newtonian (PN) theory;
IMRPhenomD uses the TaylorF2 model \citep{Buonanno:2009zt, Arun:2004hn} which is known up to 3.5PN order.
To model the intermediate and MR regions, IMRPhenomD (and all waveforms in the IMRPhenom family)
uses a series of parameterizations\footnote{
	The parameterizations for both the amplitude and phase functions can be found in \citep{khan2016frequency}.
} 
which depend purely on $\Lambda$ and can be calibrated to numerical relativity (NR) simulations.
The three sections are then \textit{stitched} together using step functions.
Importantly, the parameterizations are chosen such that they can be made $\mathcal{C}^1$ continuous at the
boundary between each section.

\begin{figure}
	\script{loss.py}
	\centering
	\includegraphics[width=\columnwidth]{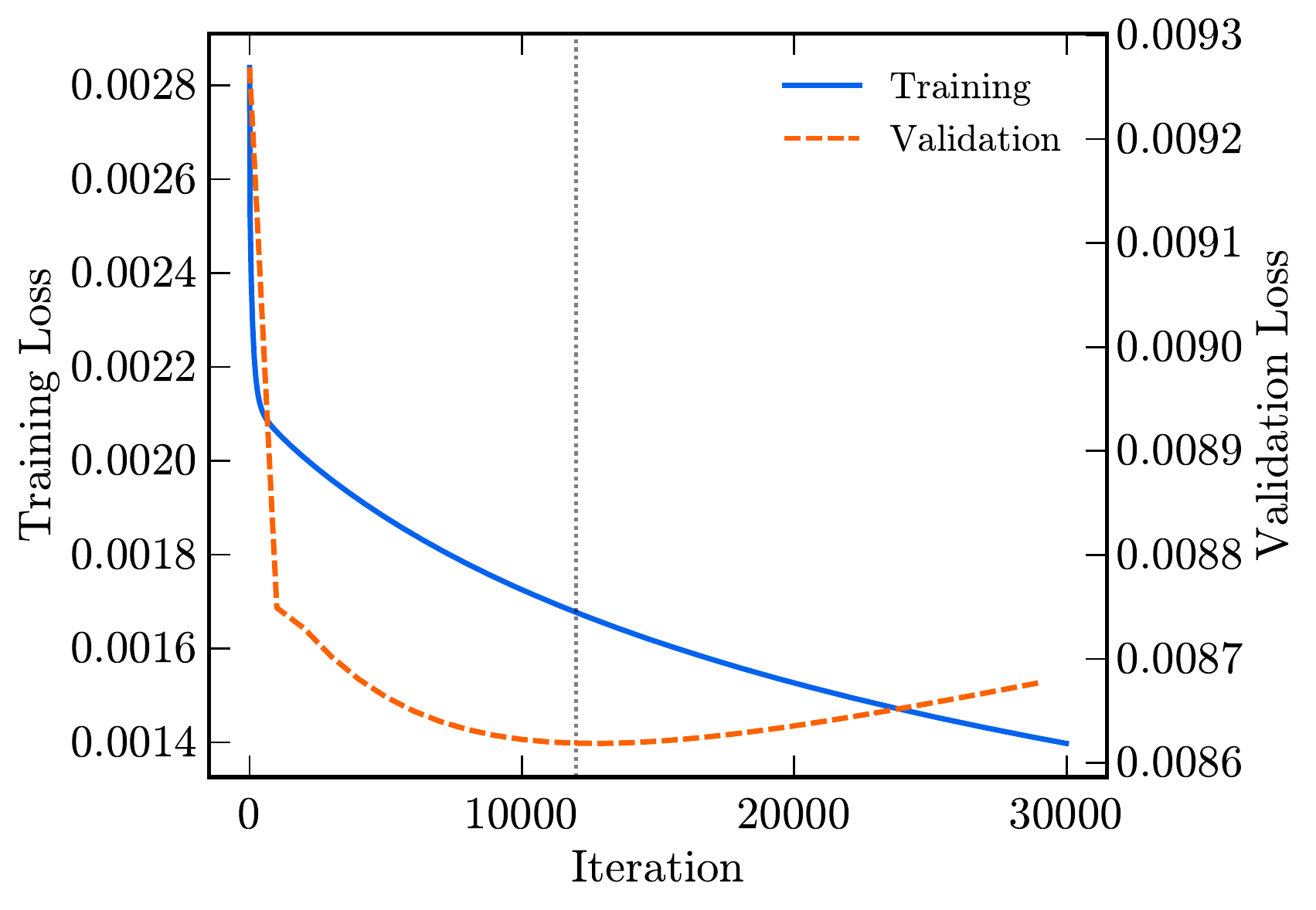}
	\caption{Loss functions against the number of iterations. The vertical
	dotted line indicates the minimum of the validation loss, which is where we stop
	the optimization.}
	\label{fig:loss}
\end{figure}

In practice the $\Lambda$ parameters are fit for each section independently, i.e., intermediate coefficients are fit whilst ignoring the MR region.
Finally, to map the grid of tuned $\Lambda$ parameters back to the intrinsic parameter space, IMRPhenomD uses the polynomial function:
\begin{align} \label{eq:Lambda}
	\Lambda^i&=\lambda_{00}^i+\lambda_{10}^i\eta \nonumber \\
	&+(\chi_{\mathrm{PN}}-1)(\lambda_{01}^i+\lambda_{11}^i\eta+\lambda_{21}^i\eta^2) \nonumber \\ 
	&+(\chi_{\mathrm{PN}}-1)^2(\lambda_{02}^i+\lambda_{12}^i\eta+\lambda_{22}^i\eta^2) \nonumber \\
	&+(\chi_{\mathrm{PN}}-1)^3(\lambda_{03}^i+\lambda_{13}^i\eta+\lambda_{23}^i\eta^2)\,,
\end{align}
where the $\lambda$'s are the fitting coefficients we are going to optimize below, $\eta$ is
the symmetric mass ratio, and $\chi_{\mathrm{PN}}$ is the post-Newtonian spin
parameter, which is defined as 
\begin{align}
	\chi_{\mathrm{PN}}=\frac{m_1\chi_1+m_2\chi_2}{m_1+m_2}-\frac{38\eta}{113}(\chi_1+\chi_2)\,.
\end{align}
Here, $m_{1,2}$ and $\chi_{1,2}$ are the primary and secondary mass and spin,
respectively. 

Although initially independent, the stitching procedure means that each section
of the waveform intrinsically depends on the full set of $\lambda$'s. 
A slightly inaccurate set of $\lambda$'s can therefore lead to inaccuracies in
the generated waveforms. 
Thus, the calibration of these coefficients is crucial to the accuracy
of IMRPhenom GW models. 
Importantly, since the fitting was performed on the individual segments,
the final waveform is not guaranteed to have $\lambda$'s close to global minima.

At the time of construction this piece-wise approach was necessary since
$\lambda$ has 209 components, making the fitting to NR simulations computationally prohibitive.
Here, for the first time we recalibrate the $\lambda$ coefficients jointly. 
This is made possible by the use of gradient-based optimization algorithms,
enabled by AD from \jax\, and {\ripple}, which are significantly more efficient in high dimensions.  

\subsection{Loss Function} \label{subsec:loss}

In order to optimize the coefficients, we need to define a loss function that
quantifies the difference between the waveform model and the target NR simulations which we want to match.
Here we adopt a quantity commonly used in GW physics called the \textit{mismatch} function \citep{owen1996search, husa2016frequency}. 
It is defined as
\begin{align} \label{eq:mismatch}
	\mathcal{M}(h_1, h_2)=1-\max_{t_0, \phi_0}\langle \hat{h}_1, \hat{h}_2\rangle,
\end{align}
where $h_{1,2}$ are the two GW waveforms we are comparing, and $t_0$ and $\phi_0$
are a relative time and phase shift respectively. 
The inner product, $\langle h_1, h_2 \rangle$, is defined as 
\begin{align}\label{eq:inner_prod}
	\langle h_1, h_2 \rangle = 4\Re\int_{f_{\mathrm{min}}}^{f_{\mathrm{max}}}\frac{h_1(f)h_2^{\ast}(f)}{S_n(f)}\,df,
\end{align}
where $\hat{h}=h/\sqrt{\langle h, h \rangle}$ is the normalized GW strain,
$S_n(f)$ is the power spectral density (PSD), and $f_{\mathrm{max}}$ ($f_{\mathrm{min}}$) is
the maximum (minimum) frequency for the integration.
We note here that the mismatch can be viewed as the mean square error (MSE) between the
two waveforms.

\begin{figure}[t]
	\script{intrin_space.py}
	\centering
	\includegraphics[width=\columnwidth]{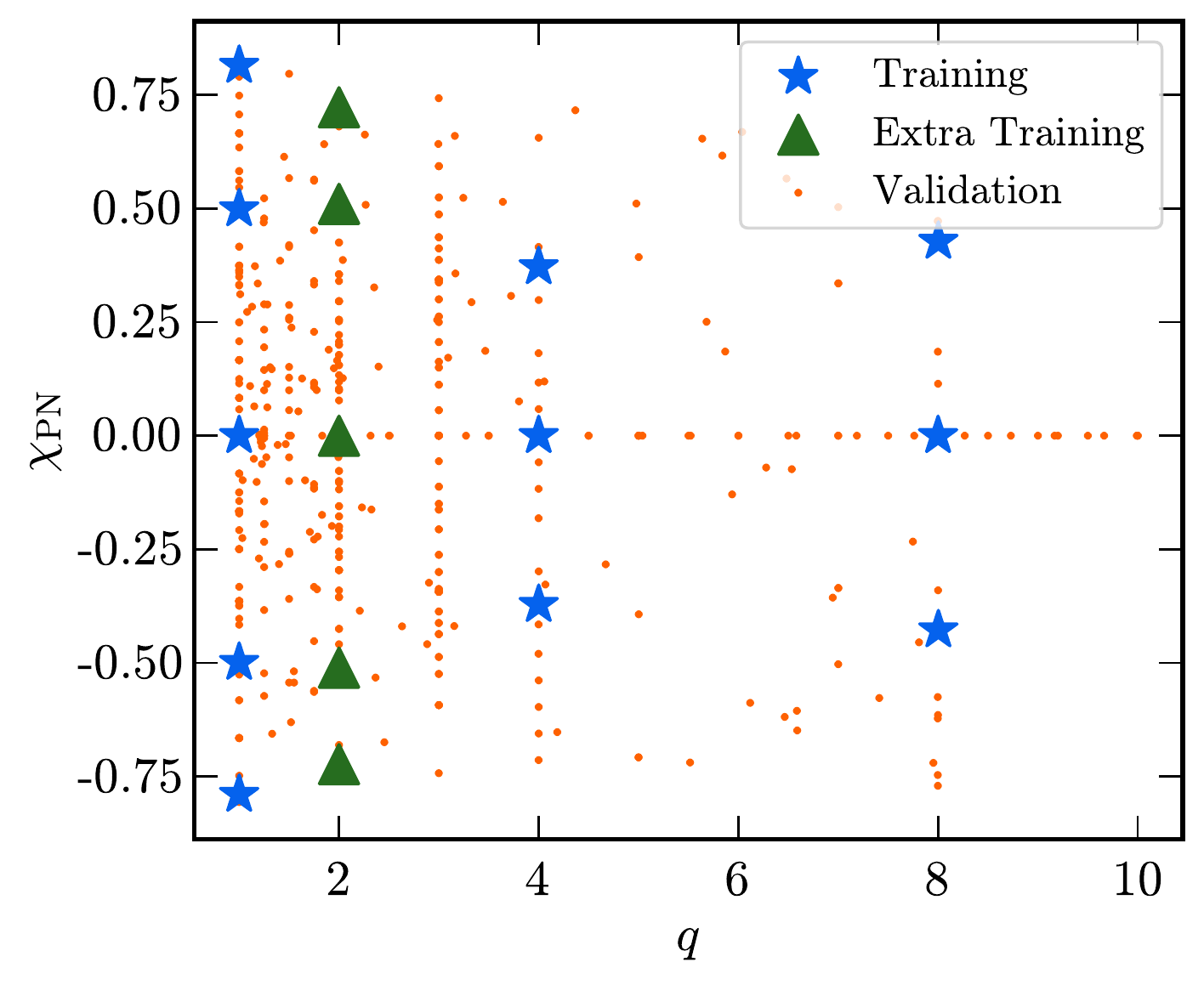}
	\caption{Distribution of training and validation NR waveforms in the
	$q-\chi_{\rm PN}$ space. The blue stars represent the waveforms used to compute
	the training loss, green triangles represent extra training data, and the 
	orange dots represent the waveforms used for validation.}
	\label{fig:intrin_space}
\end{figure}

Since we wish to optimize the model over the whole parameter space,
we need to compare multiple model generated waveforms with NR waveforms.
However, the mismatch is only defined for two input waveforms at a particular set of intrinsic parameters.
We therefore define the loss function as an average of training waveforms in two ways,
the simple average of mismatches and the normalized average of mismatches,  
\begin{align}\label{eqn:loss}
	\mathcal{L}_{\mathrm{ave}}&=\frac{1}{N}\sum_{i=1}^N\mathcal{M}_i \\
	\mathcal{L}_{\mathrm{norm}}&=\frac{1}{N}\sum_{i=1}^N\frac{\mathcal{M}_i}{\mathcal{M}_{i,\mathrm{ini}}},
\end{align}	
where $\mathcal{M}_i$ represents the mismatch of an individual training waveform,
$\mathcal{M}_{i,\mathrm{ini}}$ represents the initial mismatch of the individual
training waveform, and $N$ is the total number of training waveforms.

The two loss functions are chosen to give different 
behavior during the optimization process.
The simple average, $\mathcal{L}_{\mathrm{mean}}$, serves as the simplest choice of loss function,
but is prone to be dominated by a single point in parameter space with a large mismatch.
Other points with smaller mismatches would be insignificant comparatively, and might not be able 
to improve under such a loss function.
Alternatively, the normalized average, $\mathcal{L}_{\mathrm{norm}}$, eliminates the aforementioned issue
by encouraging the waveform to improve at each training point at a similar rate.
The ratio in $\mathcal{L}_{\mathrm{norm}}$ will therefore remain approximately the same for each training point.
Conversely, $\mathcal{L}_{\mathrm{mean}}$ allows the loss function to automatically adjust
and preferentially optimize the largest mismatches, encouraging the waveform to have similar mismatches everywhere.
In this paper, we show the results of using both loss functions 
and examine the differences between them. 

\subsection{Optimization Scheme} \label{subsec:optimization}

To compute the loss functions, we have to choose NR waveforms for calculating the mismatch.
Originally, 19 NR simulations were used to calibrate IMRPhenomD \citep{khan2016frequency}; nine of these 
are from the SXS catalog \citep{boyle2019sxs} and 10 BAM simulations. 
As BAM waveforms are not publicly available, we cannot use a training set identical to the original work.
Instead, we take the same nine waveforms from the SXS catalog plus two additional waveforms which closely mimic
two of the low mass ratio BAM simulations. 
The remaining BAM simulations have no close SXS counterparts, and are therefore not included in the training set.
Our main results therefore uses 11 NR waveforms for calibration which are listed in Tab.~\ref{tab:q148}.
Additional NR waveforms which are used for further calibration are listed in Tab.~\ref{tab:q1248}.  

The training set has a maximum mass ratio of eight due to the lack 
of high mass ratio simulations in the SXS catalog.
In fact, the SXS catalog only has NR waveforms with $q\leq10$.
Nevertheless, we are interested in the behavior of the IMRPhenomD model with small $q$,
as most BBH events observed by LIGO and Virgo have $q\leq8$.

The SXS NR waveforms are given as time-series strain.
Since IMRPhenomD is modeled in the frequency domain, we need to Fourier transform the simulation results
in order to compute the mismatch, \eqref{eq:mismatch}.
For this, we taper the time-series using Tukey window \citep{usman2016pycbc}
\footnote{
	Specifically, we choose $\alpha=2t_{\mathrm{RD}}/T$, where
	$t_{\mathrm{RD}}$ is the duration of ringdown (maximum amplitude to the end of the strain) and $T$ is the duration of the
	entire GW strain. 
} 
before using standard FFT routines to compute the fast Fourier transform.

In addition to choosing the NR waveforms for the training set,
one needs to choose a relevant PSD for the mismatch.
We have opted to use a flat PSD for the mismatch calculation,
as it provides results that are independent of the detector sensitivity and mass scale.
The use of a flat PSD ensures that the improvement in accuracy is due mainly
to the difference in high-dimensional fitting.
Additionally, we are interested in examining the effect of introducing a detector PSD
on the optimization process. For this, we have chosen the
zero-detuned high-power (\zdethp) noise PSD \citep{aasi2015advanced}.
Since the total mass of the system scales with the frequency of the waveform,
we must choose a corresponding mass scale to match the frequency range of our noise PSD.
To demonstrate the effect of introducing a detector specific PSD,
we selected an arbitrary mass scale of $M=50M_{\odot}$, as binaries
of this mass are commonly observed by the LIGO and Virgo detectors~\citep{gwtc1, gwtc2, gwtc21, gwtc3}.

We point out that our treatment of NR waveforms is different from that of \citep{husa2016frequency, khan2016frequency}.
In the original calibration process, the training waveforms are hybrids of NR and SpinAlignedEOB (SEOB) waveforms.
The low frequency inspiral part is taken from the SEOB
waveforms while the rest is taken from NR simulations.
Instead, we solely use NR waveforms for calibration since most NR waveforms used (for both training and validation)
have long enough time series data, i.e.~$>15$ orbits \citep{boyle2019sxs}, to contain part of the inspiral segment and all
merger and ringdown frequency information. We use the frequency limits
$f_{\mathrm{min}}=0.1f_{\mathrm{RD}}$ and $f_{\mathrm{max}}=1.2f_{\mathrm{RD}}$,	
where $f_{\mathrm{RD}}$ is the frequency at ringdown. This range covers most of
the IMRPhenomD's frequency range, except the minimum frequency is set slightly higher
than in the original calibration due to the NR simulation length.
When compared with IMRPhenomC, the frequency range is slightly extended to have a higher maximum
frequency \citep{santamaria2010matching}.
We use the dimensionless frequency spacing $M\Delta f=2.5\times10^{-6}$, 
which is sufficient to capture all features of the GW strain. 

With the loss function evaluated, we apply gradient descent to optimize the tunable
coefficients as shown in Algorithm~\ref{alg:gradient}. We take $\lambda_i$ to be the 
original coefficients given in \citep{khan2016frequency} because they likely lie in the
neighborhood of the minimum that we wish to find. 
We fix our learning rate, $\alpha$, to be $10^{-6}$, which is small enough to ensure we 
don't move far from the minimum.
Finally, we stop the optimization when the validation loss stops decreasing 
(see Sec.~\ref{sec:result} for description of the validation set).
This can be seen in Fig.~\ref{fig:loss} at around $12000$ iterations.

\begin{algorithm}[t]
	\caption{Gradient descent pseudocode}
	\label{alg:gradient}
	\KwIn{initial coefficients $\lambda_i$} \Parameters{number of iterations
		$N$, learning rate $\alpha$} \Variables{current coefficients $\lambda$,
		mismatch gradient $\nabla\mathcal{L}$} \KwResult{output coefficients
		$\lambda$} $\lambda\leftarrow\lambda_i$\\
	\tcc{Gradient Descent}
	\For{$i<N$}{ $\mathcal{L}\leftarrow Mismatch(\lambda)$ \\
		$\nabla\mathcal{L}\leftarrow AutoDiff(\mathcal{L})$\\
		$\lambda\leftarrow\lambda-\alpha\nabla\mathcal{L}$\\
	} \Return{$\lambda$}
\end{algorithm}

\begin{table}[t]
	\centering
	\begin{tabularx}{0.8\columnwidth}{@{\extracolsep{\fill}}lrrr}
		\toprule\midrule Code         & $q$ & $\chi_1$ & $\chi_2$ \\
		\midrule\midrule SXS:BBH:0156 & 1.0 & -0.95    & -0.95    \\
		SXS:BBH:0151 & 1.0 & -0.60    & -0.60    \\
		SXS:BBH:0001 & 1.0 &  0.00    &  0.00    \\
		SXS:BBH:0152 & 1.0 &  0.60    &  0.60    \\
		SXS:BBH:0172 & 1.0 &  0.98    &  0.98    \\
		SXS:BBH:1418 & 4.0 & -0.40    & -0.50    \\
		SXS:BBH:0167 & 4.0 &  0.00    &  0.00    \\
		SXS:BBH:1417 & 4.0 &  0.40    &  0.50    \\
		SXS:BBH:0064 & 8.0 & -0.50    & -0.46    \\
		SXS:BBH:0063 & 8.0 &  0.00    &  0.00    \\
		SXS:BBH:0065 & 8.0 &  0.50    &  0.46    \\ \midrule\bottomrule
	\end{tabularx}
	\caption{List of NR waveforms used to recalibrate the model. The mass ratio is defined as
	$q=m_1/m_2\geq 1$ and the spins are denoted by $\chi_{1,2}$. Out of the 11 waveforms listed
	here, two SXS waveforms are analogues to two of the BAM NR simulations used in the original calibration (SXS:BBH:1417 as A7 and SXS:BBH:1418 as A9 in \citep{khan2016frequency}). The rest of the SXS waveforms are also used in the original IMRPhenomD calibration.}
	\label{tab:q148}
\end{table}
\begin{table}[t]
	\centering
	\begin{tabularx}{0.8\columnwidth}{@{\extracolsep{\fill}}lrrr}
		\toprule\midrule Code         & $q$ & $\chi_1$ & $\chi_2$ \\
		\midrule\midrule SXS:BBH:0234 & 2.0 & -0.85    & -0.85    \\
		SXS:BBH:0235 & 2.0 & -0.60    & -0.60    \\
		SXS:BBH:0169 & 2.0 & 0.00     & 0.00     \\
		SXS:BBH:0256 & 2.0 & 0.60     & 0.60     \\
		SXS:BBH:0257 & 2.0 & 0.85     & 0.85     \\ \midrule\bottomrule
	\end{tabularx}
	\caption{Additional NR waveforms used in further recalibration.}
	\label{tab:q1248}
\end{table}

\begin{figure*}[t]
	\script{0154.py}
	\centering
	\includegraphics[width=\textwidth]{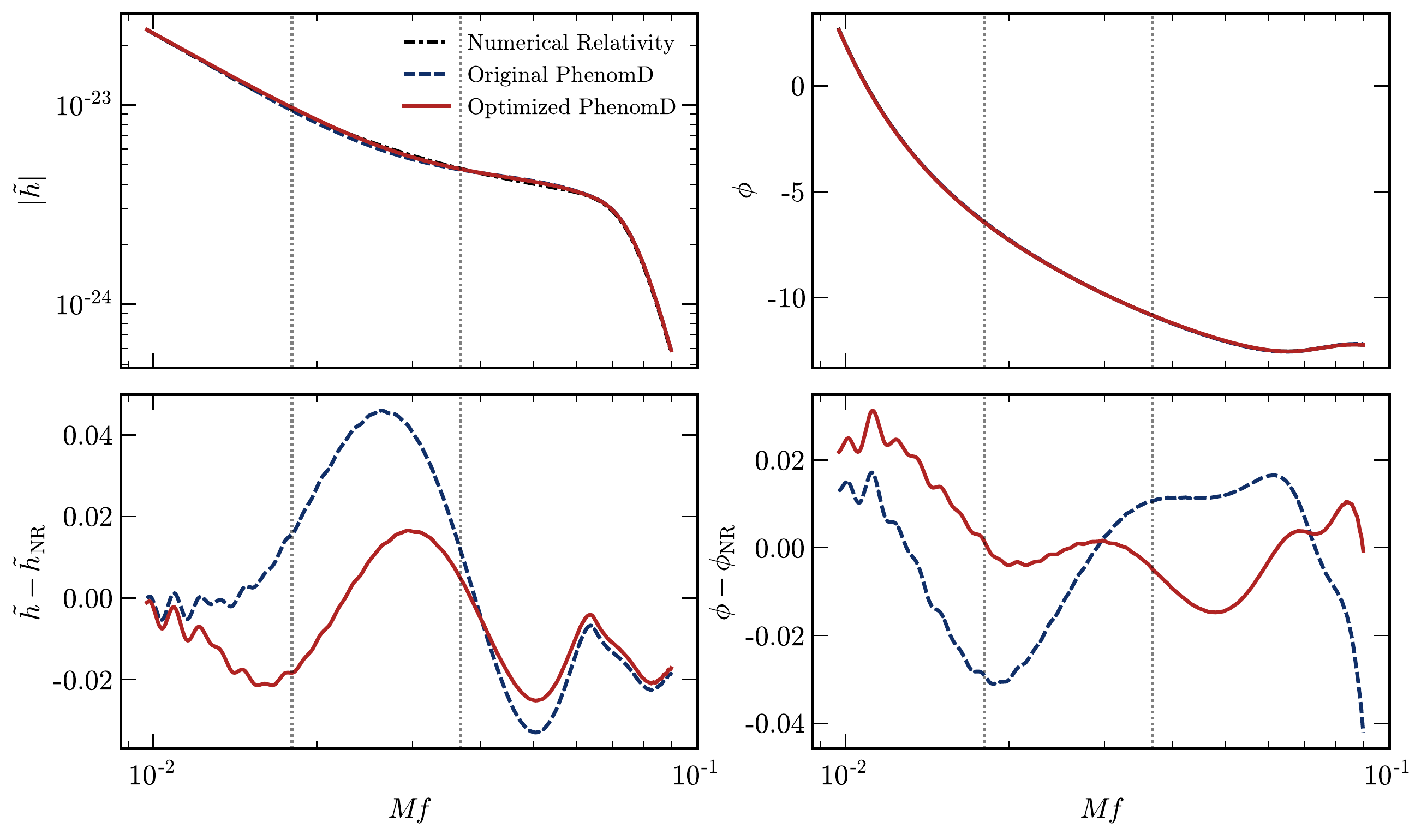}
	\caption{Comparison between original and optimized IMRPhenomD waveforms.
	Here we show the SXS:BBH:0154 NR waveform, which has a mass ratio of $q=1$ and spins
	$\chi_1=\chi_2=-0.8$. The original mismatch is $2.8\times10^{-4}$ and
	the optimized mismatch is $5.3\times10^{-5}$. \textit{Top}: Here we show the
	amplitude (left) and phase (right) of the NR, original IMRPhenomD, and optimized IMRPhenomD
	waveforms. \textit{Bottom}: Here we show the relative error between the NR and
	IMRPhenomD waveform ampltiudes (left) as well as the absolute error of the phases between the NR and
	IMRPhenomD waveforms (right).}
	\label{fig:0154}
\end{figure*}

\section{Result and Comparison with Original Model} \label{sec:result}

To evaluate how well the optimization procedure generalizes to waveforms that are
not in the training set, we evaluate the mismatch between the fine-tuned model
and an additional 526 NR waveforms in the SXS catalog i.e., the validation set.
We select waveforms that share the same part of the parameter space with the training
set, i.e., waveforms with negligible eccentricity (${e<2\times10^{-3}}$)
and precession (${\chi_{x,y}<5\times10^{-3}}$). Figure~\ref{fig:intrin_space}
shows how the training and validation waveforms are distributed in the
$q-\chi_{\rm PN}$ space.

To illustrate the effect of optimization on an individual waveform level, in Fig.
\ref{fig:0154} we plot the phase and amplitude of a particular waveform before and after optimization together with
the NR waveform taken directly from the SXS catalog. In the bottom panel one can see that,
compared to the original IMRPhenomD waveform, the optimized waveform has smaller residuals both in amplitude and
phase, particularly in the inspiral region where the amplitude displays a
$50\%$ reduction in error. For a fair comparison, we selected the SXS:BBH:0154 NR waveform
which was also used in \citep{khan2016frequency} to validate the original waveform.

\begin{figure}[t]
	\script{q148.py}
	\centering
	\includegraphics[width=\columnwidth]{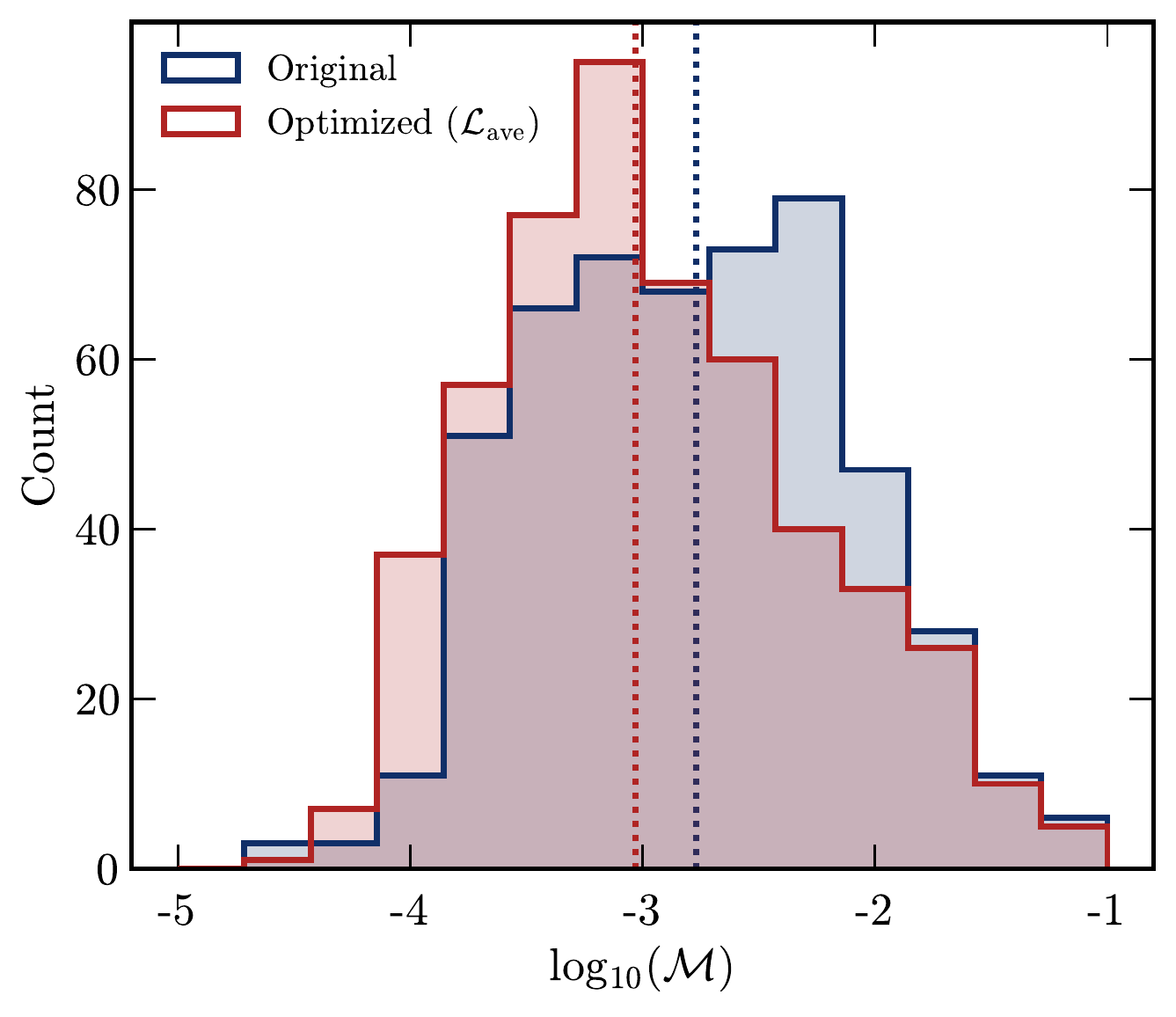}
	\caption{Distributions of mismatches before and after optimization using the 
	$\mathcal{L}_{\mathrm{ave}}$ loss function. Mismatches are calculated using the training waveforms listed in 
	Tab.~\ref{tab:q148} and	are weighted with a constant PSD. 
	The dotted lines represent the median of the distributions, which decreased by 45.3\% during optimization.}
	\label{fig:q148}
\end{figure}

With the purpose of improving downstream tasks such as parameter estimation in
mind, the more relevant metric of improvement is the distribution of improvement
in mismatch over the entire validation dataset. Figure~\ref{fig:q148} shows the
distribution of log-mismatches for the validation waveforms before and after the
optimization procedure. Here we show results using a constant PSD in our loss
function. One can see the distribution of the optimized waveform is skewed toward lower mismatches,
with the peak of the distribution being shifted by approximately an order of
magnitude, and the median mismatch being reduced by 50\% (see vertical dotted lines). When using
$\mathcal{L}_{\mathrm{norm}}$, we observe a less pronounced improvement with a 22.9\%
decrease in the median. 

\begin{figure}[t]
	\script{q148_q1248_compare.py}
	\centering
	\includegraphics[width=\columnwidth]{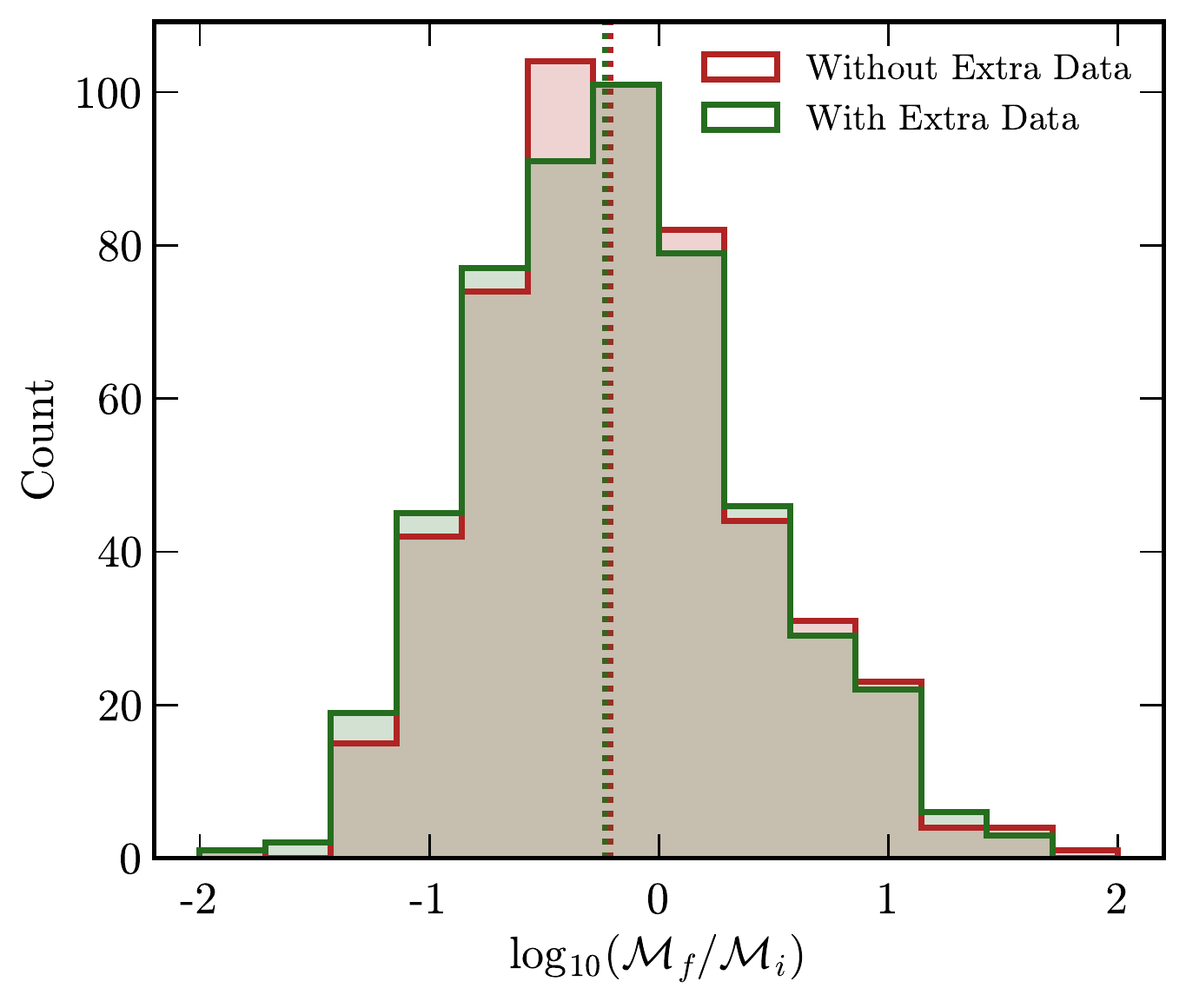}
	\caption{Distributions of $\log_{10}$ difference in mismatch. The red
	distribution uses the training waveforms listed in
	Tab.~\ref{tab:q148} while the green distribution uses waveforms
	listed in Tab.~\ref{tab:q148}~and~\ref{tab:q1248}. Mismatches are calculated
	using a constant PSD with the loss function $\mathcal{L}_{\mathrm{ave}}$. 
	The dotted lines represent the median of the distributions, which decreased by an additional 10.8\% during optimization with the additional data.}
	\label{fig:q148_q1248_compare}
\end{figure}

Note that the performance of the IMRPhenomD model was initially tested using the
{\zdethp} weighted mismatch.\footnote{
	Note, however, that the mismatch was never directly used during the calibration process~\citep{khan2016frequency, khan2019phenomenological}.
} We would therefore like to examine whether using the {\zdethp} PSD in our loss function
could lead to an improved mismatch. Performing the optimization again, we find no significant difference
between the results using the two PSDs in the distribution of mismatches.

To understand whether additional training data can further improve the
performance of the model, we include waveforms that are not present in the
original IMRPhenomD calibration in our training dataset; the parameters
can be found in Tab.~\ref{tab:q1248}. We specifically choose to use $q=2$ events
since we have abundant $q=2$ NR waveforms to validate the final result. The new
set of coefficients generated from this optimization process yields only
marginal improvements in the newly produced waveforms, as seen in Fig.
\ref{fig:q148_q1248_compare}. The high mismatch tail of the newly optimized
distribution remains comparable to the distribution from the first optimization, indicating that
the original dataset is sufficient for this task. Similarly, utilizing the
{\zdethp} PSD in the loss function together with additional waveforms results in a
similar level of improvement.

\begin{figure}[t]
	\script{ps_q148_qchi.py}
	\centering
	\includegraphics[width=\columnwidth]{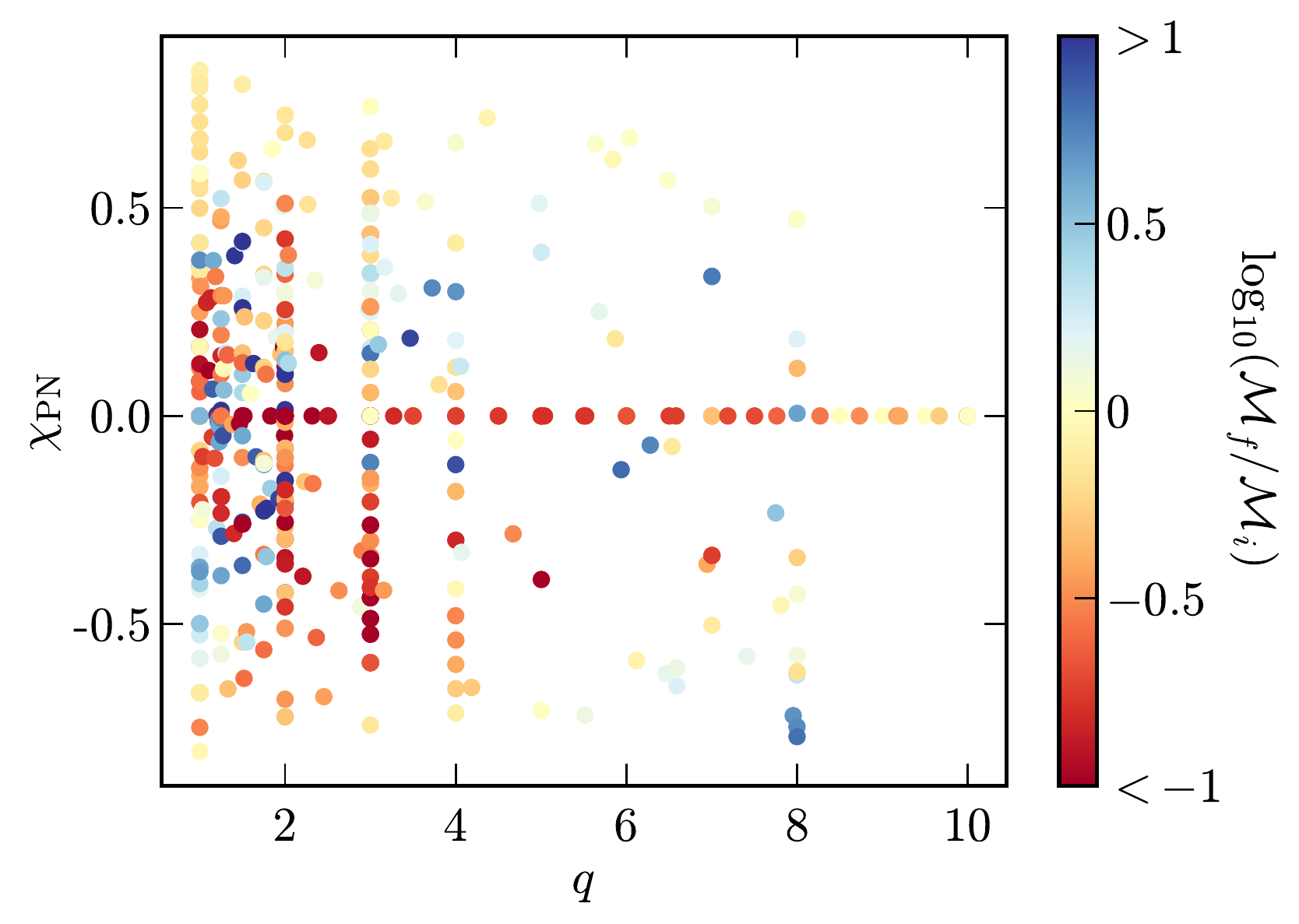}
	\caption{Fractional mismatch change of the validation waveforms in the $q-\chi_{\mathrm{PN}}$ plane. 
	We show results for the $\mathcal{L}_{\mathrm{ave}}$ loss function with a constant PSD 
	and training waveforms in Tab.~\ref{tab:q148}. Here, the colorbar represents the 
	$\log_{10}$ difference between optimized and original mismatches. Red points indicate that the waveform
	is improved by the optimization procedure, while blue points indicate that the waveform mismatch
	\textit{increases} during optimization.}
	\label{fig:ps_q148_qchi}
\end{figure}

\begin{figure}[t]
	\script{ps_q148_chi1chi2.py}
	\centering
	\includegraphics[width=\columnwidth]{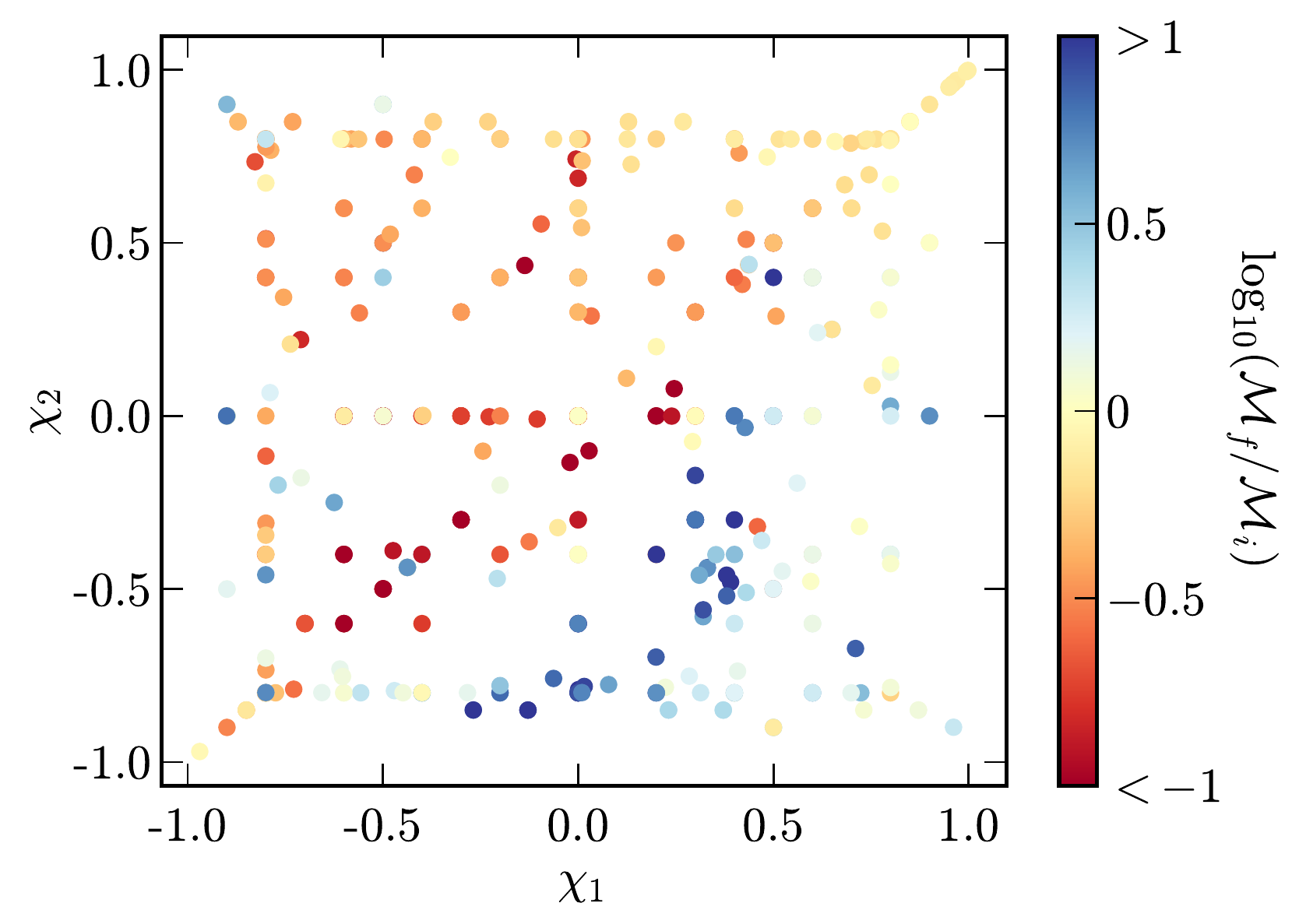}
	\caption{Fractional mismatch change of the validation waveforms in the $\chi_1-\chi_2$ plane. We
	show results for the $\mathcal{L}_{\mathrm{ave}}$ loss function with a
	constant PSD and training waveforms listed in Tab.~\ref{tab:q148}.}
	\label{fig:ps_q148}
\end{figure}

To investigate the performance of recalibration over the source parameter space,
we plot the improvement of the log-mismatch as a function of the parameter space
$q-\chi_{\mathrm{PN}}$ in Fig. \ref{fig:ps_q148_qchi}. Red points indicate that the waveform
is improved by the optimization procedure, while blue points indicate that the waveform mismatch
\textit{increases} during optimization.
We can see that waveforms with $q\leq4$ show the most consistent average improvement.
This is likely due to the better coverage of training waveforms in that part of the parameter space (see Fig.~\ref{fig:intrin_space}).
On the spin axis, we can see that the waveforms with
$\chi_{\textrm{PN}}\sim0$ show the most consistent improvement. 
When we move away from $\chi_{\textrm{PN}}\sim0$, the improvement fluctuates but exhibits an overall trend. 
This is particularly true in the $q\leq4$ region, where we see a consistent
improvement of the waveform for $\chi_{\textrm{PN}}<0$. 
We also plot the parameter space $\chi_1 - \chi_2$ in Fig.~\ref{fig:ps_q148}. 
Points along the diagonal axis, $\chi_1\sim\chi_2$, show good mismatch improvements as discussed
above. Meanwhile, the top-left and bottom-right regions respond to the optimization differently. In the top-left
region, the waveform generally improves after optimization. However,
in the bottom-right region, the waveform does not improve after optimization. 

Given that the waveform model's ansatz may not be entirely compatible with NR,
and the optimization procedure is carried out over a distribution of waveforms
with varying source parameters, it is conceivable that different parts of the
source parameter space may not share the same set of optimal IMRPhenomD
parameters.
This would mean that there are trade-offs in accuracy between different
parts of the parameter space. If this is the cause of the lack of improvement in
the high mismatch tail of the distribution, segmenting the parameter space into
smaller subspaces should alleviate this problem. On the other hand, if the
ansatz lacks the correct parameterized form to capture the NR waveforms'
behavior as a function of the source parameters, the results will always be
biased, and we should not expect any improvement, even if we segment the
parameter space during training.


\begin{table}[t]
	\centering
	\begin{tabularx}{0.8\columnwidth}{@{\extracolsep{\fill}}lrrr}
		\toprule\midrule Code         & $q$ & $\chi_1$ & $\chi_2$ \\
		\midrule\midrule SXS:BBH:0172 & 1.0 & 0.98     & 0.98     \\
		SXS:BBH:0152 & 1.0 & 0.60     & 0.60     \\
		SXS:BBH:0001 & 1.0 & 0.00     & 0.00     \\
		SXS:BBH:1417 & 4.0 & 0.40     & 0.50     \\
		SXS:BBH:0167 & 4.0 & 0.00     & 0.00     \\
		SXS:BBH:1426 & 8.0 & 0.48     & 0.75     \\
		SXS:BBH:0063 & 8.0 & 0.00     & 0.00     \\ \midrule 
		SXS:BBH:0370 & 1.0 & -0.20    & 0.40     \\
		SXS:BBH:2092 & 1.0 & -0.50    & 0.50     \\
		SXS:BBH:0330 & 1.0 & -0.80    & 0.80     \\
		SXS:BBH:2116 & 2.0 & -0.30    & 0.30     \\
		SXS:BBH:2111 & 2.0 & -0.60    & 0.60     \\
		SXS:BBH:0335 & 2.0 & -0.80    & 0.80     \\
		SXS:BBH:0263 & 3.0 & -0.60    & 0.60     \\
		SXS:BBH:2133 & 3.0 & -0.73    & 0.85     \\
		SXS:BBH:0263 & 4.0 & -0.80    & 0.80     \\ \midrule 
		SXS:BBH:0156 & 1.0 & -0.95    & -0.95    \\
		SXS:BBH:0151 & 1.0 & -0.60    & -0.60    \\
		SXS:BBH:0001 & 1.0 & 0.00     & 0.00     \\
		SXS:BBH:1418 & 4.0 & -0.40    & -0.50    \\
		SXS:BBH:0167 & 4.0 & 0.00     & 0.00     \\
		SXS:BBH:1419 & 8.0 & -0.80    & -0.80    \\
		SXS:BBH:0063 & 8.0 &  0.00    &  0.00    \\ \midrule 
		SXS:BBH:0304 & 1.0 & 0.50     & -0.50    \\
		SXS:BBH:0327 & 1.0 & 0.80     & -0.80    \\
		SXS:BBH:2123 & 2.0 & 0.30     & -0.30    \\
		SXS:BBH:2128 & 2.0 & 0.60     & -0.60    \\
		SXS:BBH:2132 & 2.0 & 0.87     & -0.85    \\
		SXS:BBH:2153 & 3.0 & 0.30     & -0.30    \\
		SXS:BBH:0045 & 3.0 & 0.50     & -0.50    \\
		SXS:BBH:0292 & 3.0 & 0.73     & -0.85    \\ \midrule\bottomrule
	\end{tabularx}
	\caption{List of NR waveforms used in recalibrating the coefficients in the four $\chi_1-\chi_2$ regions.
	From top to bottom the lines denote the top-right ($\chi_1,\chi_2>0$), top-left ($\chi_1<0<\chi_2$),
	bottom-left ($\chi_1,\chi_2<0$), and bottom-right ($\chi_1>0>\chi_2$) regions respectively.
	Note that for the top-right and bottom-left regions, waveforms are chosen to have $\chi_1\approx\chi_2$,
	while the training waveforms for the other two regions are chosen to have $\chi_1\approx-\chi_2$.}
	\label{tab:quadrants}
\end{table}

\begin{figure*}[t]
	\script{all_quadrants.py}
	\centering
	\includegraphics[width=\textwidth]{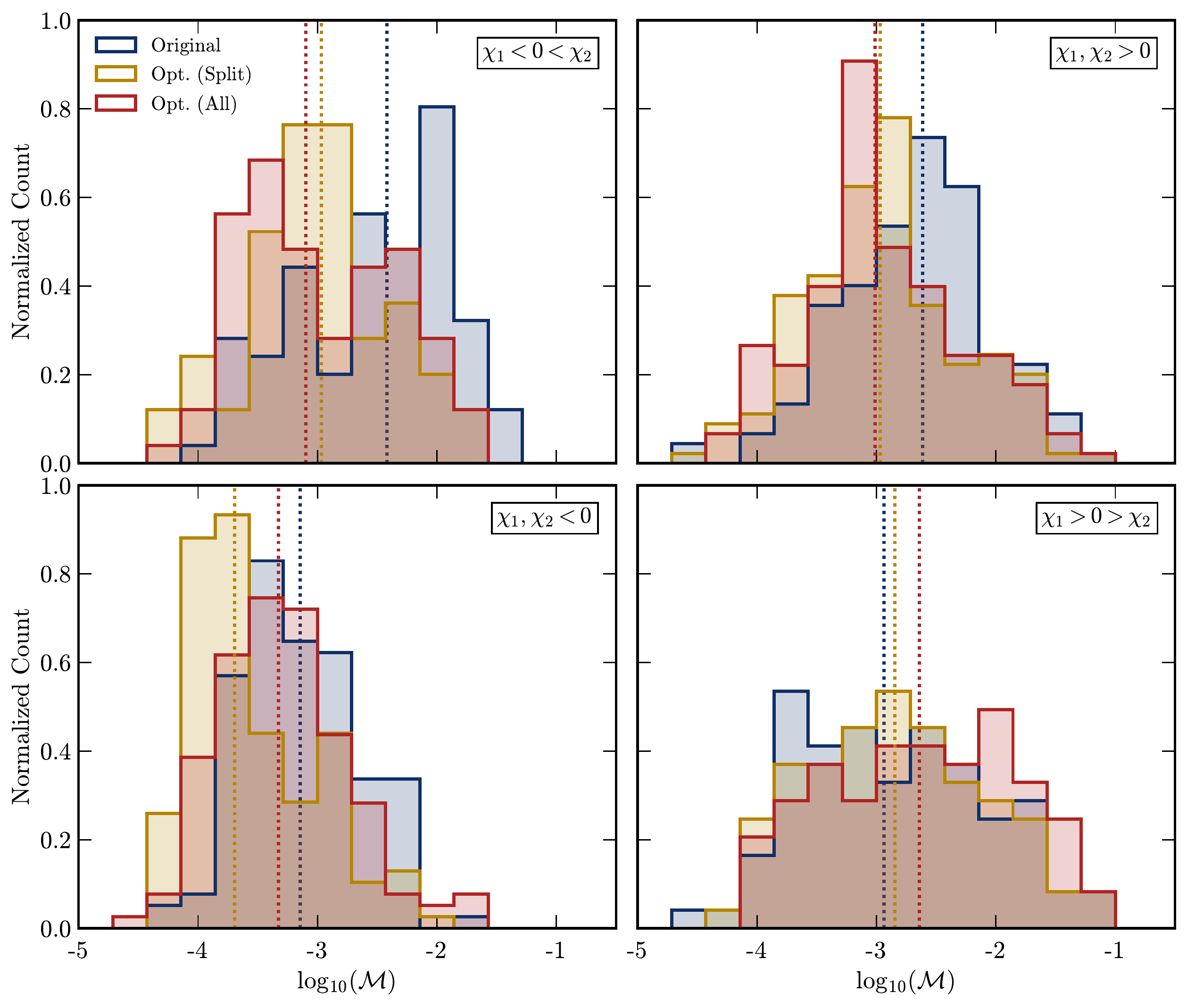}
	\caption{Distributions of mismatches for both split region optimization and all region optimization in all 4 
	regions. We use a constant PSD to calculate the mismatch and
	$\mathcal{L}_{\mathrm{ave}}$ as the loss function. The medians of each distribution are indicated by dotted
	lines.}
	\label{fig:all_quadrants}
\end{figure*}

\begin{figure}[t]
	\script{ps_q148_quadrants.py}
	\centering
	\includegraphics[width=\columnwidth]{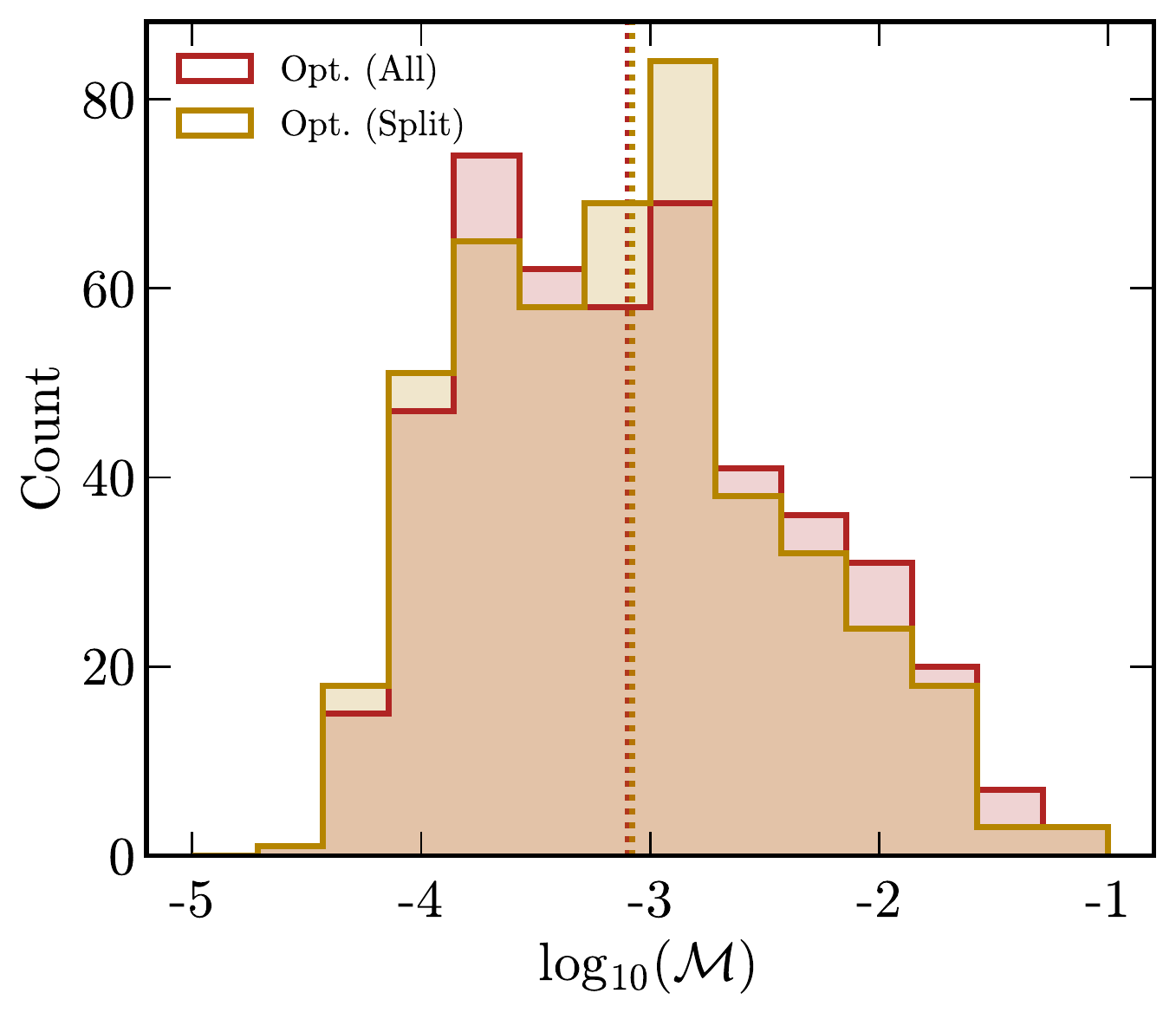}
	\caption{Combined mismatch distributions for both optimization schemes shown in Fig.~\ref{fig:all_quadrants}. The medians of both distributions are indicated by dotted lines. }
	\label{fig:ps_q148_quadrant}
\end{figure}

We divided the parameter space into four regions to analyze the effect of the
recalibration procedure on each region separately
(Fig.~\ref{fig:all_quadrants}~and~\ref{fig:ps_q148_quadrant}). The training waveforms used for fitting in
this scenario are listed in Tab.~\ref{tab:quadrants} and the loss functions 
are calculated using a simple average of the mismatches, $\mathcal{L}_{\mathrm{ave}}$. The top-left and
bottom-right regions have limited data for $q>4$, hence the result is only valid up to $q\leq4$ and we only use waveforms with $q\leq4$ as validation waveforms for these two regions. 
From Fig.~\ref{fig:all_quadrants}, we observe that both the top-right and bottom-left regions improve significantly over the original model.
This is especially pronounced for the bottom-left region, where the improvement is significantly better than optimizing all regions simultaneously.
This suggests the ansatz fits this part of the parameter space well.
The top-left region also improves over the original model although it is similar to when optimizing all waveforms at once.
On the other hand, the bottom-right region does not improve over the original model, indicating that a change to the ansatz is required to fit the NR data better.

Overall, the improvement for both optimization schemes are similar (Fig.~\ref{fig:ps_q148_quadrant}).
Although split region optimization uses more training waveforms, some of the waveforms,
e.g. opposite spins, hinders the optimization procedure and hence gives a slightly worse result when comparing with all region optimization.

\section{Discussion and Conclusion} \label{sec:discussion}

The results in this work show a promising way to understand and improve the
accuracy of phenomenological waveform models by jointly optimizing all
coefficients at once. However, there are a number of caveats as well
as multiple ways to improve upon the current work.

While this study focuses on the IMRPhenomD model due to its availability as an
automatically differentiable waveform in \ripple, other more sophisticated waveforms such as
IMRPhenomP \citep{hannam2014simple, khan2019phenomenological} and IMRPhenomXAS
\citep{pratten2020setting,pratten2021computationally} can also potentially benefit
from recalibration. In fact, since these waveforms have a larger number of calibration
parameters and are fit to a larger number of NR simulations, it is possible that the 
improvement will be more dramatic than for IMRPhenomD. {\ripple} is being regularly
updated with new waveforms, and future work should carefully examine whether these more
modern waveforms can be improved. 

The correlation we observe between the mismatches and the waveform's parameters
can also be used to design more physical ansatz. For example, we see the
recalibration process struggles to improve in the regime where $\chi_{1}$ is
positive and $\chi_{2}$ is negative. This means there is a dependency between
the two spins that is not captured by the current ansatz. A systematic scheme that
incorporates our calibration method with waveform design will be handy in
constructing new waveform models in the future. 

In making Fig.~\ref{fig:ps_q148_quadrant}, we split the parameter space into four regions purely for simplicity.
These simple cuts demonstrate the accuracy of the waveform can be further improved,
but they are almost certainly not the optimal way to incorporate the extra information
we have about the waveform. A more general way such as adding new functional dependency
between the parameters should be explored in the future.

In this work, we use the mismatch to quantify the accuracy of the waveform model. 
A natural extension of this work is to investigate how the recalibration process affects
downstream analyses such as parameter estimation and population modeling.
Differentiable samplers such as~\citep{Wong:2022xvh}, which have been recently
introduced to the community, potentially allow one to optimize the waveform directly
using metrics from parameter estimation. For example, minimizing the bias in an injection-recovery run
could be used as a loss function. Overall, this approach could help reduce systematic waveform error
in parameter estimation simply through recalibrated waveforms. We plan to investigate these avenues in
the future.

The results presented in Fig.~\ref{fig:q148_q1248_compare} indicate that
increasing the number of training NR waveforms used in the waveform optimization yields
only a marginal increase in accuracy. This observation suggests that the parameterized ansatz employed 
in IMRPhenomD struggles to capture the full complexity of the NR waveforms. 
However, as the accuracy requirements of waveform models increases with more sensitive detectors,
more NR waveforms will be required to train more flexible Phenom models. 
Accurate NR simulations must therefore be developed in parallel with waveform models
to ensure we meet future detector accuracy requirements.

Overall, the development of accurate waveform models is crucial for the success of GW astronomy.
In this work, we have explored how modern computational tools (automatic differentiation and gradient descent)
can help with this task. 
Our method is general and can be applied to any waveform model that is differentiable.
We therefore encourage the waveform development community to utilize these tools
and hope that this work can effectively contribute to the development of accurate waveform models.

\section{ACKNOWLEDGMENTS}

We thank Will M.~Farr, Max Isi, and Mark Hannam for helpful discussions; we also
thank Carl-Johan Haster, Neil J.~Cornish and Thomas Dent for comments on the
draft. The Flatiron Institute is a division of the Simons Foundation. T.E.\ is
supported by the Horizon Postdoctoral Fellowship.

\bibliography{bib}

\end{document}